# Optimized Multi-Contrast Self-Supervised MRI Reconstruction using Learned k-space Partitioning

Brenden Kadota, *Member*, *IEEE*, Charles Millard, and Mark Chiew, *Member, IEEE*

***Abstract*— Objective: Deep Learning has shown promise in accelerating MRI by reconstructing high-quality images from under-sampled data. While recent work has leveraged multi-contrast information to improve reconstruction performance, these methods rely on supervised learning, which requires fully sampled k-space for training. One method, self-supervised learning via data undersampling (SSDU), enables direct training on under-sampled k-space by partitioning it into two sets, with a network mapping between the two. In this work, we improve MRI self-supervised MRI reconstruction with two modifications. Methods: We propose a multi-contrast self-supervised learning framework that jointly trains on multiple under-sampled contrasts without requiring fully sampled k-space data as a reference. Moreover, we learn an optimal self-supervised data partitioning for each contrast in an end-to-end manner, further enhancing reconstruction quality. Specifically, we learn an optimal partitioning probability distribution, which is sampled to generate a mask for partitioning. Results: Experiments on two publicly available multi-contrast MRI datasets demonstrate the improved reconstruction quality of our proposed self-supervised multi-contrast learned partitioning method compared to the current single-contrast self-supervised learning methods. We also demonstrate that learning the partitioning of k-space data further enhances the fidelity of reconstructions. Conclusion: Multi-contrast reconstruction combined with learned partitioning improves reconstruction fidelity over single-contrast self-supervised MRI reconstructions. Significance: Our method can facilitate higher image fidelity and/or accelerated MRI protocol times compared to previous self-supervised methods, and without requiring fully sampled k-space for training.**

## I. INTRODUCTION

Magnetic resonance imaging (MRI) clinical protocols routinely acquire multiple contrasts from the same anatomical region to provide complementary information for clinicians. Contrasts such as T1-weighted, T2-weighted, and fluid-attenuated inversion recovery (FLAIR) provide clinicians with critical information for more effective patient management. However, the sequential acquisition of contrasts, combined with the inherently slow process of acquiring the Fourier representation of an image, known as k-space, results in lengthy clinical MRI protocols. One common approach to accelerating MRI scans is to acquire fewer k-space measurements, a strategy typically applied to various extents across all scans in a protocol. However, reconstructing sub-Nyquist Fourier data requires advanced reconstruction techniques. Parallel imaging uses the linear independence of spatial receiver sensitivities to resolve missing values [1], [2], whereas compressed sensing exploits image compressibility and incoherent sampling to reconstruct images from under-sampled data [3].

As multiple contrasts from the same anatomy often contain highly correlated information, several studies have explored combining multiple contrasts in reconstruction methods to enhance reconstruction performance. Two approaches have emerged for multi-contrast reconstruction: (1) using a fully sampled contrast as auxiliary information in reconstruction [4], [5], [6], [7], [8], [9] and (2) jointly reconstructing multiple under-sampled contrasts. Given that clinical protocols often under-sample all scans, we focus on the latter. Multi-contrast parallel imaging methods have been explored that use multiple contrasts as additional virtual coils [4]. Compressed sensing methods, such as Bayesian compressed sensing [10], joint sparsity [11], [12] and patched based on sparsity [13], [14], have been explored to leverage the correlated sparsity between contrasts.

With the success of deep learning (DL) in MRI reconstruction [15], [16] DL multi-contrast reconstructions have also gained popularity. These methods typically use a single neural network to jointly reconstruct multiple contrasts with the network implicitly learning inter-contrast relationships. A common strategy of combining contrast information for deep learning reconstruction is to concatenate contrasts along the channel dimension [6], [17], [18], [19] and passing to a convolutional neural network. Alternatively, some approaches leverage the transformer attention mechanisms [20] to transfer contrast information more effectively, [21], [22].

This paragraph of the first footnote will contain the date on which you submitted your paper for review. It will also contain support information, including sponsor and financial support acknowledgment. For example, "This work was supported in part by the U.S. National Science Foundation under Grant BS123456". Brenden Kadota and Mark Chiew are with Department of Medical Biophysics, University of Toronto, Toronto, On, Canada and Physical Sciences Research Platform, Sunnybrook Research Institute (e-mail: Brenden.kadota@mail.utoronto.ca, mark.chiew@utoronto.ca ).

Charles Millard., was with Welcome Centre for Integrative Neuroimaging, FMRIB, University of Oxford, Oxford (e-mail: charliemillard@live.co.uk).

Multi-contrast networks can effectively utilize multiple contrasts to outperform single-contrast networks of the same architecture.

While multi-contrast methods focus on improving reconstruction by exploiting inter-contrast relationships, another critical factor in achieving high-quality results is the under-sampled k-space acquisition locations. Optimizing the k-space acquisition pattern has been explored in compressed sensing reconstructions. Studies have used methods such as hand-picked probability distributions [23], [24], greedy algorithms [25], and Bayesian optimization [26], to optimize k-space sampling locations for compressed sensing. Additionally, learned sampling strategies for multi-contrast reconstruction have been investigated in both compressed sensing and parallel imaging, with studies finding that segregated sampling across contrasts yields superior performance [27], [28].

Deep learning methods have also been used to learn the optimal k-space sampling patterns in conjunction with the reconstruction network. Two methods of deep learning k-space sampling pattern optimization have been explored: parameterization of k-space locations and learned probabilistic methods.

Parameterization of k-space converts k-space locations into a learnable parameter that can be optimized through backpropagation and gradient descent. Learning a parameterized k-space sampling pattern is an extremely non-convex problem and difficult to optimize. Additionally, parameterizing k-space transforms the reconstruction problem into non-cartesian because the k-space parameters are continuous. Multiple parameterization methods have been explored using maximum likelihood [29], line constraints [30], neural ordinary differential equations [31], and b-spline parameterization [32]. Parameterizing k-space has also been combined with multiple contrasts in [33], where optimal k-space sampling over multiple contrasts was used to further improve performance.

The second deep learning optimized sampling approach involves learning a probability distribution from which k-space samples are drawn [34], [35], [36]. Since sampling is a stochastic operation that backpropagation gradients cannot pass through, the reparameterization trick [37] is used to enable gradient-based optimization of the probability distribution. As relaxation techniques provide gradients only at the transition zone, relaxations such as the straight-through method enables the calculation of meaningful gradients [38]. Notably, the "learning optimized under-sampling probability distribution" (LOUPE) [34] strategy is an end-to-end learned under-sampling pattern method that relaxes sampling to a sigmoid function. Studies building upon LOUPE have incorporated multi-contrast information, which can help guide k-space sampling locations across multiple contrasts [28], [33].

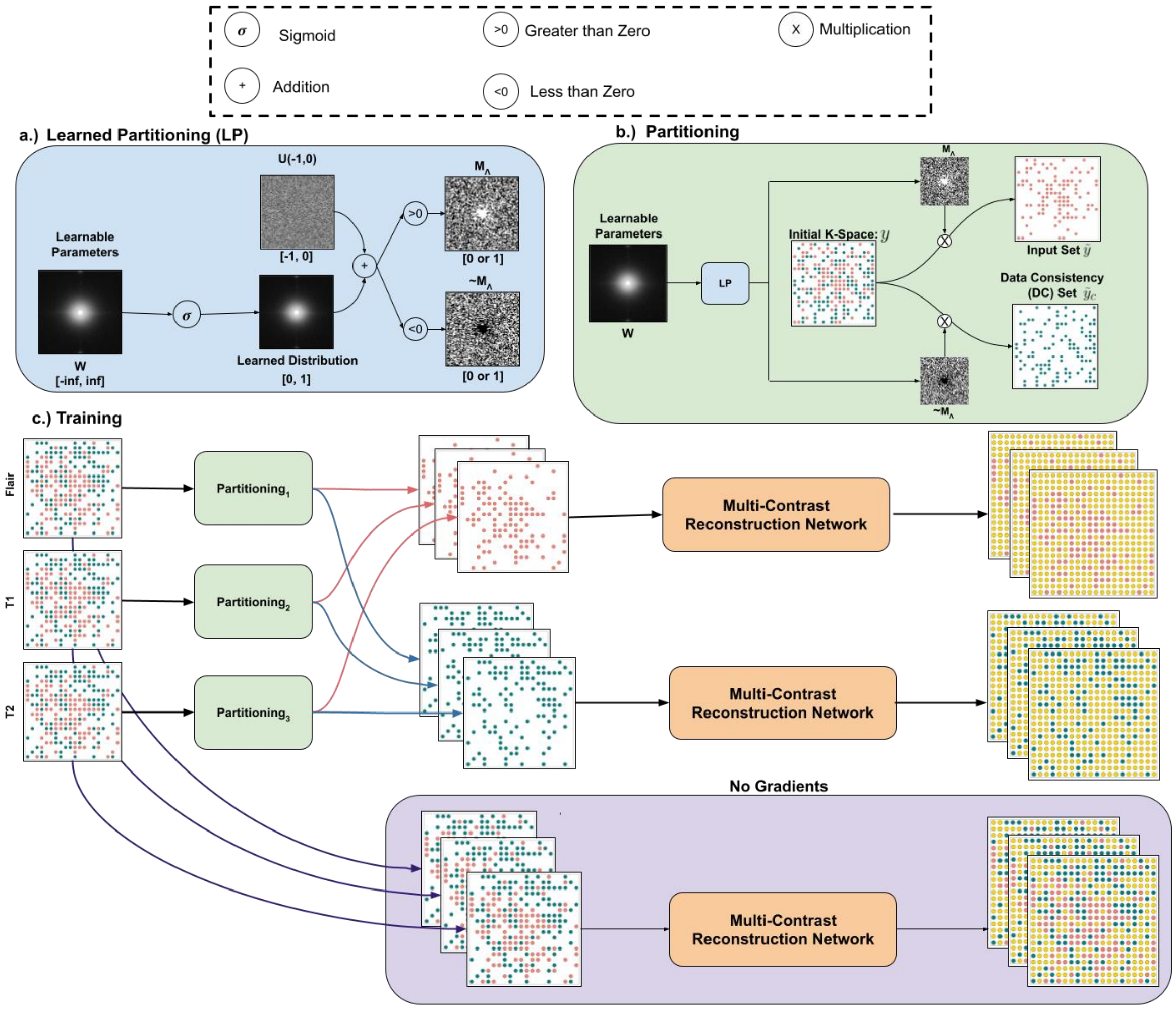

FIG. 1. a.) Learned partitioning module for learning optimal partitioning distribution. The learned probability distribution is sampled by adding noise and thresholding b.) Partitioning module that partitions under-sampled k-space into two sets based on M and ~M. c.) Training process. Each contrast is passed through an independent partitioning module. Resulting multi-contrast sets are concatenated along the channel dimension before passing to the reconstruction network. The bottom path with un-partitioned data does not calculate gradients.

The majority of previously mentioned deep-learning studies relied on supervised learning for training, which requires fully sampled k-space for computing the loss function. However, in many cases acquiring fully sampled k-space may not be feasible due to extended scan times and motion. Additionally, in certain scan types, such as single-shot echo-planar imaging, the T2* signal decay limits the full acquisition of k-space. To address this, self-supervised reconstruction methods have been proposed to train directly on under-sampled k-space data. Remarkably, at inference, self-supervised methods can perform comparably to supervised training without ever accessing fully sampled data.

One self-supervised method, self-supervised learning via data under-sampling (SSDU) [39], [40] partitions under-sampled k-space into two sets using a neural network that maps between the two. Improved performance in SSDU can be achieved by using a dual-domain loss function, which computes a loss in both image space and k-space. The dual domain loss function has shown significant improvement on self-supervised reconstructions [41]. Further work has improved performance by partitioning k-space by the same distribution as the initial under-sampling distribution [42], and by explicitly learning the partitioning probability distribution using deep learning. However, to the best of our knowledge, there has been no prior work on multi-contrast self-supervised reconstruction, nor on integrating it with learned partitioning, making our approach the first to explore both.

Here, we introduce a novel training framework that integrates multi-contrast information into self-supervised learning to improve reconstruction fidelity. Additionally, we optimize the partitioning distribution for each contrast in our multi-contrast self-supervised learning network to further improve reconstruction quality.

Our key contributions are as follows:

• We demonstrate that multi-contrast self-supervised learning outperforms single-contrast self-supervised learning, particularly at high acceleration factors.

• We propose a learned partitioning strategy for multi-contrast self-supervised learning, which surpasses heuristic-based partitioning approaches.

• We validate our method on a variety of under-sampling patterns and on multiple datasets.

• We additionally characterize multi-contrast self-supervised performance when varying the number of jointly reconstructed contrasts.

## II. Methods

### A. Multi-Contrast Self-Supervised Learning

The multi-contrast MRI acquisition can be modeled as a forward process:

$$y_l = M_{\Omega,l} A x_l + n_l, \quad (1)$$

where $x_l \in \mathbb{C}^n$ is the $l^{th}$ $(0 < l < L)$ contrast image, with $y_l$ being the corresponding under-sampled k-space, $A$ is the forward linear encoding model incorporating the Fourier Transform and coil sensitivities, and $n_l$ is Gaussian noise for the $l^{th}$ contrast. The undersampling mask $M_{\Omega,l}$ for the $l^{th}$ contrast is independently drawn from a probability distribution function Ω, where $\Omega \sim \prod_{i=0}^{n} \mathcal{B}(\mathrm{p}[i])$ . Ω represents a Bernoulli distribution, with each voxel is described by some independent Bernoulli distribution with probability p[i] of being sampled. SSDU trains directly on under-sampled k-space by partitioning under-sampled data into two disjoint sets and training a network to map between the two [39]. Here, we extend SSDU for multi-contrast reconstruction by independently splitting under-sampled k-space into disjoint sets. Given a mask $M_{\Lambda,l}$ for contrast $l$, we partition multi-contrast k-space by the equation:

$$\tilde{y}_l = M_{\Lambda,l} y_l \quad (2.1)$$

$$\tilde{y}_{c,l} = \sim M_{\Lambda,l} y_l, \quad (2.2)$$

where $\sim M_{\Lambda,l}$ is the complement mask of $M_{\Lambda,l}$. Typically, the partitioning distribution $\Lambda_l$ which draws $M_{\Lambda,l}$ is hand-picked and chosen heuristically. The sets $\tilde{y}_l$ and $\tilde{y}_{c,l}$ have disjoint k-space points and are referred to as the input set and data consistency set respectively.

For simplicity, we have decided to use channel concatenation to test our training framework. All contrasts are concatenated along the channel dimension $\{\tilde{y}_i\}_1^L = \tilde{Y}$ before passing through a single reconstruction network. The same k-space loss function in SSDU is calculated independently across contrasts, resulting in L additional k-space loss terms:

$$\mathcal{L}_{k_1} = \frac{1}{L} \sum_{l=1}^{L} \left\| M_{\Omega,l} \sim M_{\Lambda,l} f_\theta(\tilde{Y})_l - \tilde{y}_{c,l} \right\|_1^1, \quad (3)$$

where $f_\theta$ is a multi-contrast neural network parameterized by $\theta$. The subscript $l$ on $f_\theta(\tilde{Y})_l$ represents indexing the reconstructed k-space of the $l^{th}$contrast because $f_\theta(\tilde{Y})$ jointly reconstructs all contrasts. The masking terms $M_{\Omega,l} \sim M_{\Lambda,l}$ ensure the loss is only calculated where k-space points exist in $\tilde{y}_{c,l}$. This loss function can be seen as learning a mapping between the k-space set $\tilde{Y}$ to $\tilde{Y}_c$.

At inference, the neural network can reconstruct full multi-contrast k-space by passing the original under-sampled multi-contrast data concatenated along the channel dimension $\{y_i\}_1^L =$ Y through the network $f_\theta(Y)$ to reconstruct fully sampled multi-contrast k-space [39], [42]. We perform a final data consistency gradient step at inference by replacing estimated k-space lines with the acquired ones:

$$Y_{inference} = \sim M_\Omega f_\theta(Y) + Y, \quad (4)$$

where $\sim M_\Omega$ is the complement of masks across all contrasts.

### B. Learned Partitioning

Similar to our previous work, we learn the probability distribution that draws the partitioning masks $M_\Lambda$. To achieve this, we learn an optimal probability distribution for each

*Fig. 1.* a.) Self-supervised k-space loss calculated independently across contrasts. Both input and data consistency k-space estimates are used to calculate a k-space loss b.) Image space loss calculated independently across contrasts. $\mathcal{L}_{i_1}$ is the image loss between the input and DC recon. $\mathcal{L}_{i_2}$ is the image loss between input and initial recon. $\mathcal{L}_{i_3}$ is the image loss between DC and initial recon.

contrast. Let the partitioning probability distribution $\Lambda$ for a single contrast $l$ be defined as:

$$\sigma_t(W_l) = \Lambda_l, \tag{5}$$

where $W_l \in \mathbb{R}^n$ are learnable weights for contrast $l$; $\sigma_t$ is an element-wise sigmoid with a slope controlled by the hyperparameter $t$: $\sigma_t(W_l) = \frac{1}{1+e^{-tW_l}}$. The resulting $\Lambda_l$ is a contrast specific learned partitioning distribution which is an independent Bernoulli distribution per-voxel: $\Lambda_l = \prod_{i=0}^{n} \mathcal{B}(p_i[i])$. Because the gradients cannot easily be calculated from the sampling probability distributions, the distribution is modified to an equivalent representation using the reparameterization trick [37]. The resulting sampling equation becomes:

$$M_{\Lambda,l} = (\Lambda_l + U) > 0, \tag{6}$$

where $U \in \mathbb{R}^n$ is the uniform distribution between [-1,0] independently drawn for each voxel, and $> 0$ is an element wise threshold. Equation 6 is an equivalent representation of sampling from a Bernoulli distribution but allows gradients to pass to our learned probability distribution $\Lambda_l$. However, as the greater than zero threshold only contains non-zero gradients at the transition zone, the straight-through method is used to calculate the meaningful gradients [35], [38]. The straight-through method calculates Equation 6 in the forward pass while approximating its gradient in the backward pass using the derivative of a sigmoid function:

$$\nabla M_{\Lambda,l} = \sigma_s(Z_l)\sigma_s(Z_l - 1), \tag{7}$$

where $\sigma_s$ is sigmoid with a slope of $s$ and $Z$ is the forward pass activation $Z_l = \Lambda_l + U(-1,0)$. The whole partitioning process can be visualized in Figures 1a and 1b.

### *C. Dual Domain Loss Function*

We additionally employ a dual domain loss function to calculate the loss both in k-space and image space to improve reconstruction performance. Similar to [41], we independently pass k-space sets $Y$, $\tilde{Y}$, and $\tilde{Y}_c$ as input to the reconstruction network to get three multi-contrast reconstructions (Figure 1c). We omit gradient calculations for $f_\theta(Y)$ to speedup training. Due to the additional reconstructions of $f_\theta(Y)$, $f_\theta(\tilde{Y})$, and $f_\theta(\tilde{Y}_c)$ additional k-space loss terms can be calculated. The modified k-space loss term from equation 3 becomes:

$$\mathcal{L}_{k_1} = \frac{1}{L}\sum_{l=1}^{L} \left\| M_{\Omega,l} \sim M_{\Lambda,l} f_\theta(\tilde{Y})_l - \tilde{y}_{c,l} \right\|_1^1, \tag{8a}$$

$$\mathcal{L}_{k_2} = \frac{1}{L}\sum_{l=1}^{L} \left\| M_{\Omega,l} M_{\Lambda,l} f_\theta(\tilde{Y}_c)_l - \tilde{y}_l \right\|_1^1, \tag{8b}$$

Equation 8a is the same as Equation 3, which calculates a mapping between $\tilde{Y}$ and $\tilde{Y}_c$. The additional k-space loss term in Equation 8b can be seen as the opposite. Equation 8b calculates a mapping between $\tilde{Y}$ and $\tilde{Y}$. $f_\theta(Y)$ is omitted from the k-space loss terms as it always will evaluate to zero due to the final data consistency step at inference (Equation 4).

Additionally, to the k-space loss terms in Equation 8a+8b, we compute a pair-wise image space loss between the three reconstruction outputs, again independently across contrasts:

$$\mathcal{L}_{i_1} = \frac{1}{L}\sum_{i=1}^{L} \mathcal{L}_{img}\left(rss(F^{-1}f_\theta(\tilde{Y})),\ rss(F^{-1}f_\theta(Y))\right),$$

$$\mathcal{L}_{i_3} = \frac{1}{L}\sum_{i=1}^{L} \mathcal{L}_{img}\left(rss(F^{-1}f_\theta(\tilde{Y})),\ rss(F^{-1}f_\theta(\tilde{Y}_c))\right),$$

$$\mathcal{L}_{i_3} = \frac{1}{L}\sum_{i=1}^{L} \mathcal{L}_{img}\left(rss(F^{-1}f_\theta(\tilde{Y}_c)),\ rss(F^{-1}f_\theta(Y))\right), \tag{9}$$

where $F^{-1}$ is the inverse Fourier Transform; $rss$ is the root sum of squares operator. For $\mathcal{L}_{img}$ we use the same loss as [22] which is the L1 loss combined with the vertical and horizontal image gradients. The equation is given below:

$$\mathcal{L}_{img}(u,v) = \|u - v\|_1^1 + \|\nabla_x u - \nabla_x v\|_1^1 + \left\|\nabla_y u - \nabla_y v\right\|_1^1$$

where u, v are the two images in the loss function.

The final loss function, with 5L terms, is:

$$\mathcal{L} = \lambda\mathcal{L}_{k_1} + (1-\lambda)\mathcal{L}_{k_2} + \beta_1\mathcal{L}_{i_1} + \beta_2\mathcal{L}_{i_2} + \beta_3\mathcal{L}_{i_3} \tag{3.9}$$

where $\beta_1$, $\beta_2$, and $\beta_3$ are image space loss scaling factors; and $\lambda$ is a k-space loss scaling factor. $\lambda$ is scaled between [0, 1] to balance the importance of the $f_\theta(\tilde{Y})$ and the $f_\theta(\tilde{Y}_c)$ reconstruction output on the loss. The loss calculation is depicted in Figure 2.

### *D. Multi-Contrast Reconstruction Network*

As SSDU is compatible with any reconstruction network, we choose an end-to-end VarNet which has demonstrated

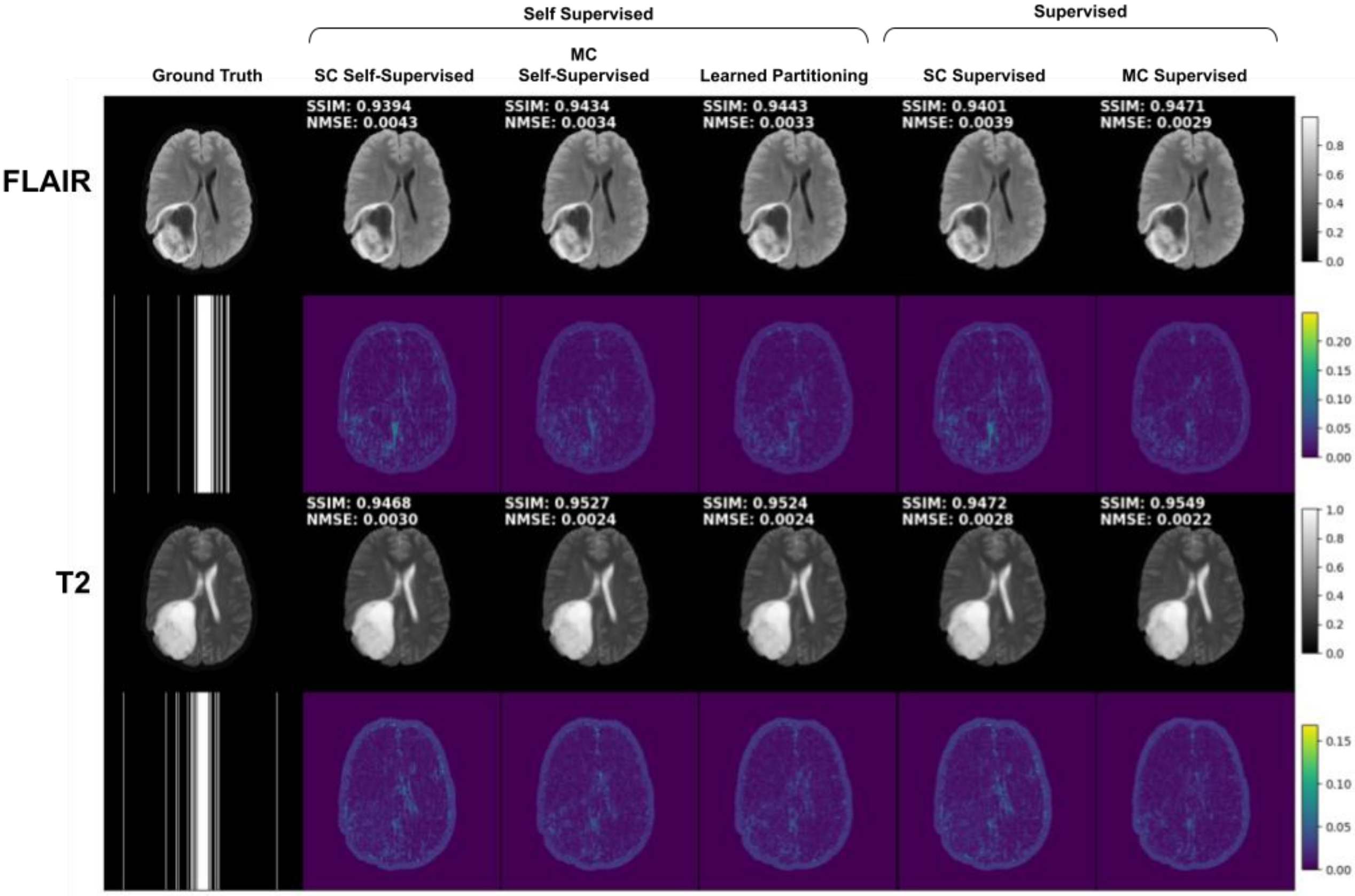

*FIG. II*. Example reconstructed slices from the 2D variable density pattern at R=8. Error maps magnified by 4 shown below. Quantitative metrics, SSIM and NMSE, shown above.

*TABLE 1*. Quantitative Metrics for Equal Space Parallel Imaging at R=8 on the BraTS dataset.

| | | | SC Self-Supervised | MC Self Supervised | MC Learned Partitioning | Supervised[*] |
|---|---|---|---|---|---|---|
| T1 | SSIM | 0.954±0.006 | 0.958±0.005 | **0.962±0.005**[†] | 0.969±0.003 | 0.980±0.005 |
| | NMSE | 0.0048±0.0005 | 0.0043±0.0004 | **0.0041±0.003**[†] | 0.0027±0.0003 | 0.0024±0.0004 |
| T2 | SSIM | 0.958±0.006 | 0.960±0.008 | **0.965±0.007**[†] | 0.972±0.006 | 0.978±0.007 |
| | NMSE | 0.0093±0.0007 | 0.0093±0.0006 | **0.0091±0.004**[†] | 0.0065±0.0004 | 0.0061±0.0005 |
| FLAIR | SSIM | 0.957±0.010 | 0.957±0.007 | **0.962±0.011**[†] | 0.968±0.006 | 0. 968±0.005 |
| | NMSE | 0.0072±0.0008 | 0.0067±0.0006 | **0.0060±0.004**[†] | 0.0045±0.0003 | 0.0045±0.0004 |
| T1CE | SSIM | 0.953±0.008 | 0.958±0.008 | **0.962±0.010**[†] | 0.977±0.006 | 0.978±0.004 |
| | NMSE | 0.0083±0.0006 | 0.0079±0.0005 | **0.0075±0.004**[†] | 0.0058±0.0003 | 0.0052±0.0005 |

*Runs marked with [*] are supervised and are trained with fully sampled k-space*
*Runs marked with [†] have a p value < 0.05 compared to other self-supervised methods.*

*TABLE 2*. Quantitative Metrics for 1D variable density at R=6 on M4Raw dataset.

| | | SC Self-Supervised | MC Self Supervised | Learned Partitioning | Supervised[*] | MC Supervised[*] |
|---|---|---|---|---|---|---|
| T1 | SSIM | 0.875±0.09 | 0.882±0.07 | **0.887±0.007**[†] | 0.879±0.009 | 0.894±0.009 |
| | NMSE | 0.0152±0.003 | 0.0145±0.002 | **0.0130±0.004**[†] | 0.0150±0.004 | 0.0123±0.003 |
| T2 | SSIM | 0.848±0.09 | 0.845±0.06 | **0.848±0.012**[†] | 0.849±0.010 | 0.859±0.011 |
| | NMSE | 0.0353±0.005 | 0.0350±0.005 | **0.0345±0.006**[†] | 0.0355±0.07 | 0.0298±0.005 |
| FLAIR | SSIM | 0.872±0.09 | 0.879±0.008 | **0.885±0.009**[†] | 0.875±0.009 | 0.895±0.008 |
| | NMSE | 0.0137±0.002 | 0.0121±0.004 | **0.0115±0.003**[†] | 0.0129±0.004 | 0.0105±0.003 |

*Runs marked with [*] are supervised and are trained with fully sampled k-space*
*Runs marked with [†] have a p value < 0.05 compared to other self-supervised methods.*

competitive performance on the fastMRI challenge [15]. The end-to-end VarNet estimates coil sensitivities while recursively enforcing k-space data consistency and refining the image space using a neural network. We modify both the data consistency and image refinement steps to accommodate for multi-contrast reconstruction.

The data consistency steps are modified to independently enforce data consistency for each contrast. The image refinement block is modified for multi-contrast images by concatenating contrast images along the channel dimension before passing into the U-Net based image refinement module. Similar to the original VarNet paper [15] the real and imaginary components of each contrast are split along the channel dimension resulting in $2L$ input channels. For the image refinement network, we use a U-Net with 64 initialfilters and 4 down-sampling layers [43]. The step size parameter η in the VarNet is modified to allow independent scaling for each contrast. The multi-contrast VarNet is unrolled for 12 cascades unless otherwise specified. Each contrast in the multi-contrast VarNet architecture introduces approximately ~15K additional parameters primarily from the initial and final filters in the U-Net.

### *E. Data*

The proposed method was evaluated on two public multi-contrast datasets: the BraTS 2019 and the low field M4Raw dataset.

*BraTS 2019*: As there are no public multi-contrast high field k-space datasets, we simulate a k-space dataset using the publicly available Brain Tumor Segmentation (BraTS) challenge 2019 dataset [44].

The BraTS dataset is a multi-site, multi-field strength and multi-contrast magnitude image dataset used for brain tumor segmentation. The dataset contains 4 contrasts: T1-weighted, T2-weighted, FLAIR, and T1 contrast enhanced from a gadolinium-based contrast agent (T1CE). As the original task was for segmentation, the data has been post-processed with skull stripping, zeroed background noise, co-registration, and interpolation to a 1mm isotropic resolution of 240x240 and 176 slices.

To simulate realistic k-space, we first intensity normalizes each axial slice to a maximum value of 1. We then apply an acquired sensitivity map which was coil-compressed to 10 channels [45]. Next, we introduce phase information into each contrast. To compute a phase map, we multiply complex white gaussian noise by k-space gaussian smoothed ($\sigma = 10mm^{-1}$) magnitude k-space data. Multiplying random noise with a smoothed k-space ensures a random phase with similar structure to the image. The resulting noise is Fourier transformed, and the resulting phase is used to generate our image phase maps. Using this phase generation method, we calculate one base phase map which is shared between all contrasts, along with a contrast specific phase map for each contrast. we weigh the base phase by 0.9 and the contrasts specific phase by 0.1 before adding the phase to the image. The resulting complex contracts with phases are converted to k-space where white, zero-mean gaussian noise with standard deviation of $\sigma = 10^{-2}$ is added for both real and imaginary components. The original dataset contains ~2000 volumes, from which we randomly selected 200, 100, 20 volumes for training, testing, and validation respectively. We select every 3rd slice for simulation from slice index 50 onwards resulting in 3200, 1600, 320 slices for training, testing, and validation.

*M4Raw*: we additionally use the M4Raw dataset [46] which is a real acquired k-space dataset from a 0.3T four-channel full-body scanner. The acquired contrasts are FLAIR, T1-weighted, and T2-weighted. All scans have a resolution of 256x256 with 18 slices and 240mmx240mm field of view. Unlike the BraTS2019 dataset, the M4Raw dataset is not co-registered between contrast of the same individual. We decided not to co-register these contrasts as small inter-contrast voxel differences have found to only modestly inhibit performance [17]. The data consists of 183 healthy volunteers with 3-5 repetitions per contrast. We split the volumes based on the pre-defined dataset splits and extract the first repetition for training. Only the first repetition was used, as phase variations between repetitions made averaging in k-space unreliable. The resulting dataset split is (2304/540/450) slices for training, testing, and validation.

### *F. Implementation Details*

We implement all models in PyTorch and train on a single NVIDIA A100 GPU with 40GB of VRAM. All training data was normalized by the mean value of the under-sampled root sum-of-squares image. we use the Adam optimizer [47] with a learning rate of $10^{-3}$ and train for 50 epochs with a batch size of 2. We set the k-space loss scaling factor λ to 0.55 and $\beta_1$ to $8x10^{-6}$, $\beta_2$ to $1x10^{-6}$ and $\beta_3$ to $5x10^{-6}$ based on a hyperparameter grid search. The center 10x10 region of k-space is left sampled for both $\tilde{y}$ and $\tilde{y}_c$ for coil sensitivity estimation using the VarNet. The VarNet is unrolled for 12 cascades for a reconstruction network with 372M parameters. The learned partitioning model adds an additional ~65K parameters per contrast. For data augmentation, we apply a random shift of [-5, 5] voxels in the coronal and sagittal planes.

### *G. Evaluation Methods*

All metrics are computed in image space after an inverse Fourier transform and root sum of squares coil combination. We additionally mask out the background before calculating the quantitative metrics, to focus on image quality within signal regions. We generate the background mask by gaussian blurring the ground truth image with a σ=11mm and taking a threshold greater than 0.1. We use two quantitative metrics for evaluating network performance, structural similarity index and normalized mean squared error. The Structural Similarity Index Measure (SSIM) is a metric between -1 and 1 with high visual similarity at 1 and complete dissimilarity at -1. we use the default parameters for SSIM with a kernel size of 7x7 based on the fastMRI benchmark [48]. We additionally use the normalized mean squared error (NMSE) which measures the relative voxel difference between images and is given by the formula $\|u - v\|_2^2 / \|u\|_2^2$ where $u$ is the reference image and $v$ is the reconstructed output.

We additionally explore sampling metrics from our learned partitioning distributions. First, we define $\tilde{R}_l = \frac{1}{n}\sum_i^n \Lambda_l[i]$ as the average probability in the learned partitioning distribution. This partitioning $\tilde{R}_l$ can be interpreted as the proportion of k-space in each partition, with the data consistency partition containing more k-space points at higher $\tilde{R}_l$ values and fewer points at low $\tilde{R}_l$ values. At $\tilde{R}_l = 2$, k-space is evenly split between sets. We additionally calculate segregated sampling index between contrasts to determine segregated sampling. In previous multi-contrast sampling work, segregated sampling, between contrast has shown to improve performance [4], [27]. To determine segregated sampling, we calculate the overlap as $O = (\sum_l^L diag(M_l)) > 1$ which counts the number of k-space locations sampled multiple times across contrasts. we calculate the percentage of overlap as:

$$\%\ overlap = \frac{\sum_i^n 1(O_i)}{n} \times 100$$

where $1(\cdot)$ is the indicator function which outputs 1 if the condition of $O_i$ is true and 0 otherwise. Segregated sampling occurs at lower overlap percentages. At a lower overlap percentage each contrast shares fewer sampled k-space locations with other contrasts, resulting in a large aggregate k-space coverage.

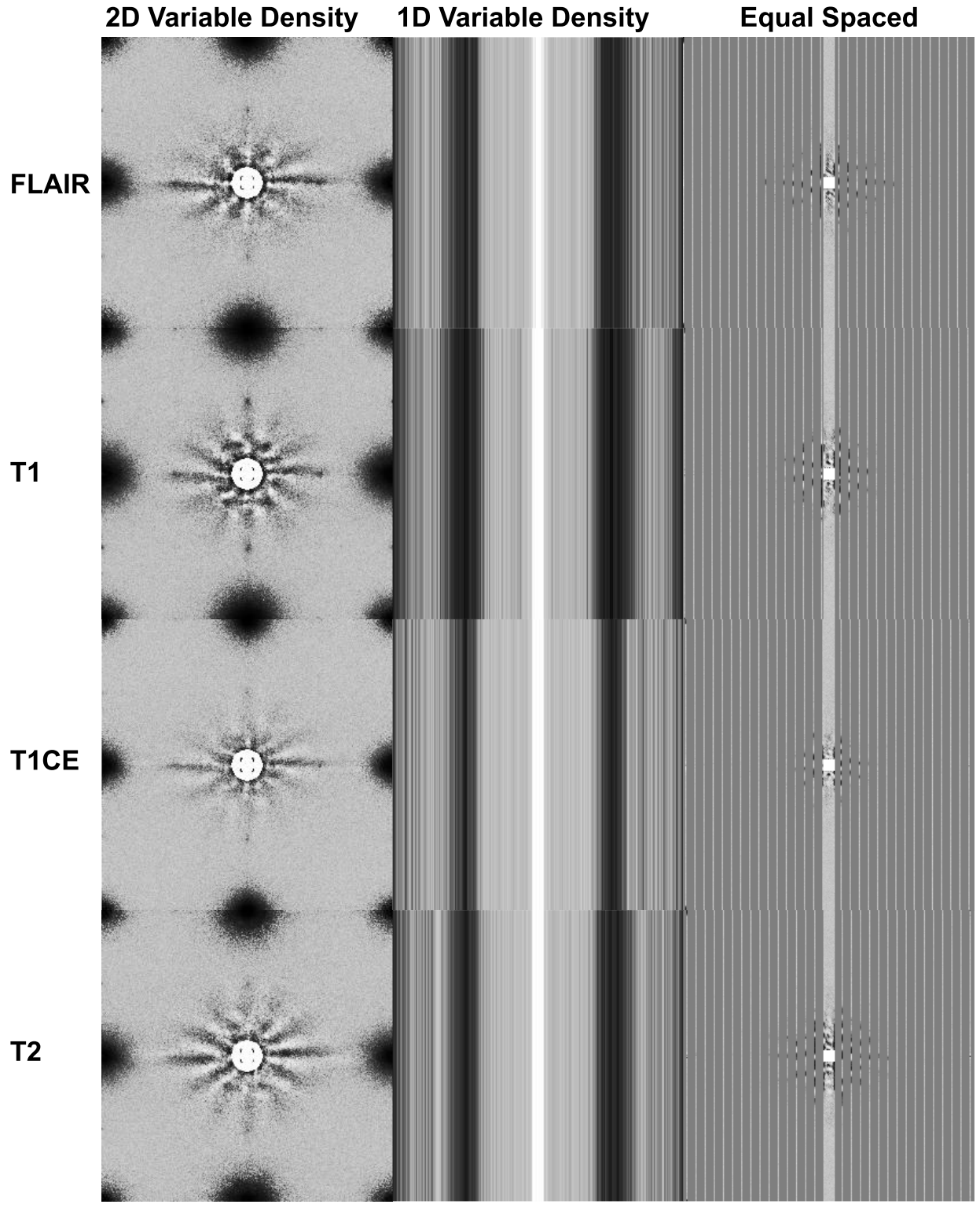

*FIG. 1.* Learned partitioning distributions for different undersampling patterns and contrasts

### *H. Comparative Methods*

Our first comparative self-supervised method is a single contrast SSDU variant that is partitioned based by the same initial under-sampling distribution as $M_\Omega$ which we call single contrast (SC) self-supervised. We additionally explore the effects of multiple contrasts on performance. We train a multi-contrast version of SSDU without learned partitioning which we call multi-contrast (MC) self-supervised. Multi-contrast self-supervised partitions k-space independently for each contrast based on the same initial under-sampling distribution. We additionally implement two fully supervised methods as baselines. Our two supervised methods are a single-contrast (SC) supervised network, and a MC supervised network which are both trained with an L1 k-space loss. All comparative methods use a VarNet architecture for reconstruction with the same number of cascades and U-Net channels. The inference time is the same between all multi-contrast comparative methods. However, the single-contrast networks are slightly faster at inference due to having ~13K fewer parameters.

## III. RESULTS

### *A. Performance on Different Initial Sampling Patterns*

Our first set of experiments explore the effects of the initial sampling distribution Ω. We tested 3 types of sampling distributions: 1D variable density, 2D variable density and equal-spaced parallel imaging under-sampling patterns at R=8. The variable density 2D and 1D distributions use a polynomial order 8 sampling distribution. Both $\tilde{y}$ and $\tilde{y}_c$ k-space sets for the 1D and equal-spaced undersampling patterns retain the center 10 phase encode lines and 2D patterns retain a center 10x10 region for coil sensitivity estimation required by the VarNet.

Example reconstructed contrasts from the BraTS dataset are shown in Figure 3 with their error maps below their reconstructed images. Here, we use the 1D variable density under-sampling pattern at an acceleration factor of R=8. Our proposed method, multi-contrast learned partitioning achieves the highest SSIM and lowest NSME of all self-supervised methods. Additionally, adding multi-contrast self-supervised models has a noticeable improvement in performance over single contrast self-supervised methods shown from the quantitative metrics and error maps. Remarkably, multi-contrast self-supervised methods surpass the performance of single-contrast supervised methods which have access to fully sampled data. MC supervised perform with lowest error and highest fidelity images and are performance baselines.

Table 1 shows the entire test metrics for all contrasts on the equal-spaced parallel imaging undersampling pattern Again, learned partitioning has the highest SSIM and lowest NMSE for both initial undersampling patterns compared to heuristically partitioned k-space. Additionally, multi-contrast self-supervised reconstruction has a considerable improvement in performance over single-contrast self-supervised methods. Notably, the performance gap between single- and multi-contrast methods is smaller for the 2D variable density distribution compared to the other undersampling patterns. Fully sampled methods perform the best with the muti-contrast methods showing the best performance in quantitative metrics.

Figure 4 presents the learned partitioning distributions for each contrast. These distributions to some extent resemble the initial under-sampling distribution which follows the theoretical justification for SSDU. Interestingly, near the center of k-space, a star like pattern emerges in partitioning probability distributions when the initial undersampling distribution is 2D and equal-spaced. Additionally, the 2D distribution has low probability regions along the edges of k-space. The 1D distribution also exhibits low-probability regions around the quarter and three-quarter positions of the k-space phase-encoding lines. All partitioning distributions converge to a partitioning acceleration factor of $\tilde{R} = 1.3$

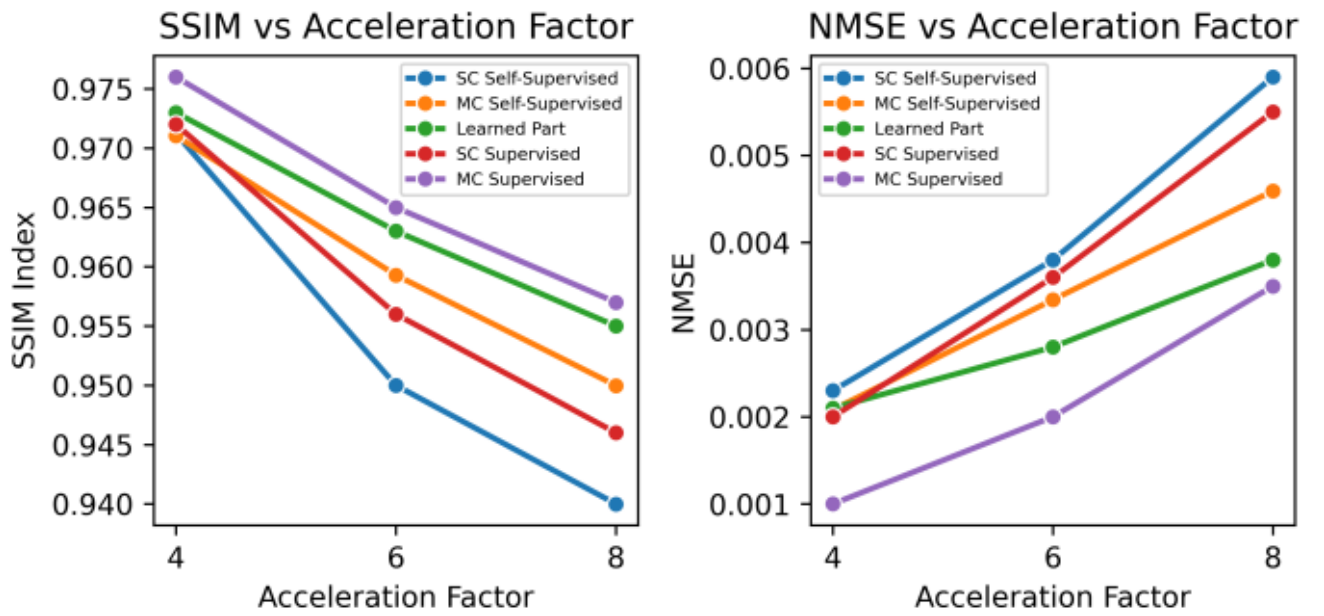

*FIG. 2.* Different reconstruction performance on the 1D variable density pattern at different acceleration factors.

To test if partitioning distributions contain segregated sampling, we calculate the percentage of k-space overlap between contrasts across 1000 draws from our learned partitioning probability distribution. We additionally calculated the percentage of k-space overlap using only the T1 partitioning distribution for all contrasts. We find there is negligible difference between the learned multi-contrast partitioning overlap at 68% and using the T1 partitioning for all contrasts at 67%. These results suggest no segregated sampling due to this minimal difference in partitioning overlap. These results are supported visually, as the learned partitioning distributions across different contrasts demonstrate high similarity.

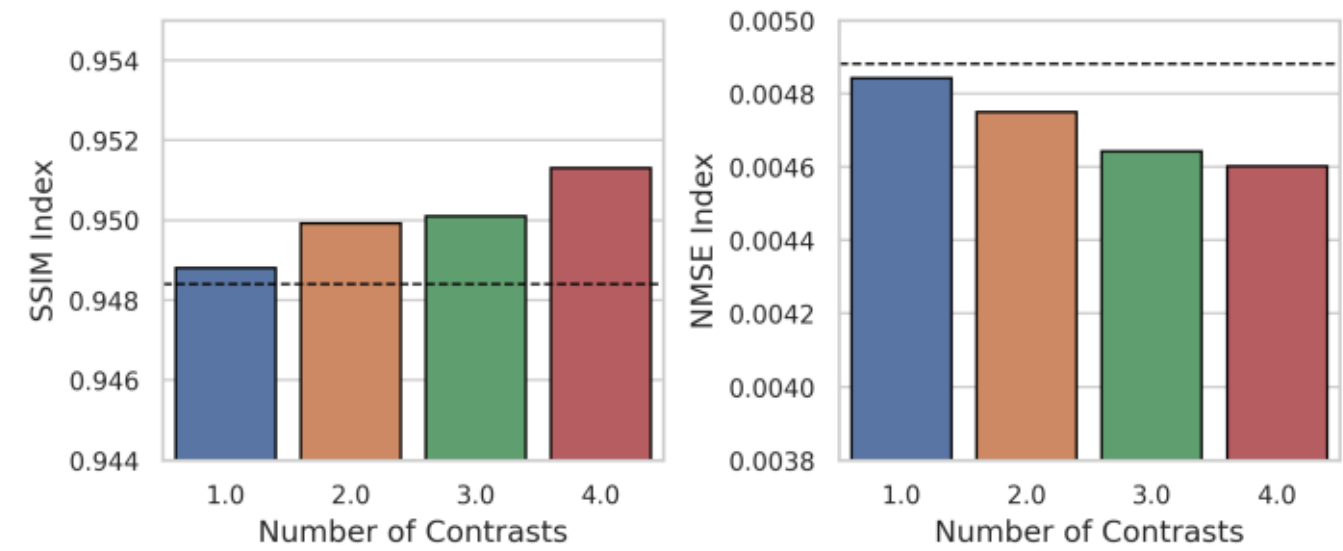

*FIG. 6.* SSIM (top) and NMSE (bottom) compared to number of contrasts. Metrics are averaged over all contrast permutations. The dashed line represents supervised single-contrast reconstruction.

### *B. Effect of Acceleration Factor*

To test our methods robustness of the reconstruction methods against different acceleration factors, we train our model at acceleration factors 4, 6, and 8 on the 1D variable density under-sampling pattern. SSIM and NMSE values were averaged across all contrasts and plotted (Figure 5). At R=4, all supervised and self-supervised learning methods perform comparably, with the supervised learning methods showing a slight improvement. At higher acceleration factors (R=6, 8) multi-contrast self-supervised learning outperforms single contrast supervised which has access to fully sampled data during training. Additionally, learned partitioning improves upon multi-contrast self-supervised performance at the higher acceleration factors. Multi-contrast self-supervised learning acts as a performance ceiling for all acceleration factors.

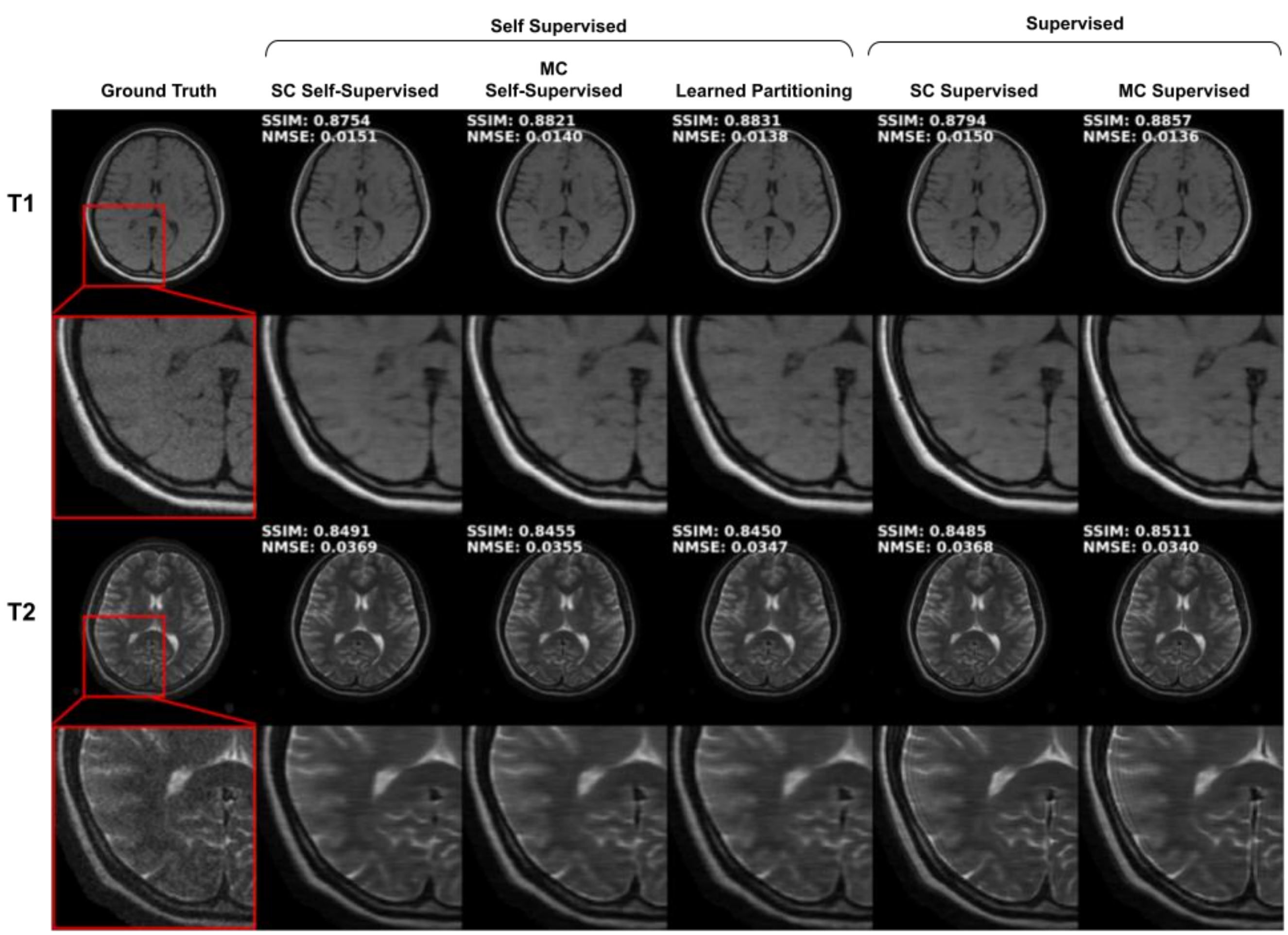

*FIG. 3.* Reconstructed slices from the M4Raw dataset on variable density R=6 1D undersampling.

*TABLE 3.* Ablation results and total training time

| | NMSE | SSIM | Training Time | Inference Time | Model Parameter |
|---|---|---|---|---|---|
| SC Self-Supervised | 0.0020 | 0.924 | 8h | 82ms | 372M |
| SC Learned Partitioning | 0.0019 | 0.924 | 10h | 101ms | 372M |
| MC Dual Domain | 0.0018 | 0.927 | 12h | 101ms | 372M |
| MC All Gradients | 0.0018 | 0.928 | 18h | 101ms | 372M |
| MC Learned Partitioning | **0.0018** | **0.928** | 12h | 101ms | 372M |

### *C. Different Number of Contrasts*

We next test the effects of the number of contrasts on reconstruction performance. We train multi-contrast learned partitioning networks for each combination of one, two, three, and four contrasts on the BraTS dataset. We use the 1d variable density undersampling pattern at an acceleration factor of R=8. We subsequently report the quantitative metrics averaged within each group of multi-contrast models, where all 2-contrast permutations are averaged together, all 3-contrast permutations are averaged together, and so on. Within each model, the metrics are further averaged across its constituent contrasts.

In Figure 6, we show SSIM and NMSE. The dashed line represents supervised single-contrast reconstruction. At two contrasts, performance of multi-contrast networks surpass single-contrast supervised metrics. As additional contrasts are added to the reconstruction network, SSIM performance increases and NMSE improve monotonically. Each contrast provides additional information to improve performance.

### *D. Performance on the M4Raw Dataset*

Figure 7 shows R=6 1D variable density under-sampling reconstructions on real acquired data on the M4Raw dataset, with Table 2 displaying the total test set metrics. All three contrasts in the dataset are plotted with the average test set metrics on the top left. We chose to show magnified regions of the reconstructed images because the error maps are noisy. All reconstructions perform well, with minimal residual aliasing artifacts. Multi-contrast learning partitioning with self-supervised learning is the best self-supervised method with sharper features. Single contrast supervised learning performs worse than both of our methods, multi-contrast self-supervised and multi-contrast self-supervised learned partitioning. Both of our methods, multi-contrast self-supervised and multi-contrast learned partitioning outperform single-contrast supervised learning which has access to fully sampled data. Multi-contrast learned partitioning has sharper features than multi-contrast self-supervised. The multi-contrast supervised baseline has the sharpest features and lowest error.

### *E. Ablations*

Finally, we empirically justify our method by performing ablations on our method on an acceleration factor of R=8 and 1D variable density sampling. We performed 3 separate ablations. The first ablation removes the non-gradient calculation for the initial data $Y$ pathway, which we call MC All Gradients. The next ablation removes the learned partitioning model and trains a dual domain self-supervised reconstruction network, which we call MC Dual Domain, which is partitioned based on the original distribution of $M_{\Omega}$. Finally, our last ablation trains a single contrast self-supervised learned partitioning method, name SC Partitioning. Table 3 shows the SSIM and NMSE of all the ablations with single contrast self-supervised and multi-contrast learned partitioning as baselines. The best ablation model is MC all gradients, but the training time is 1.5 times longer due to the additional gradients through the reconstruction network for the $y$ pathway. Dual domain multi-contrast shows a notable improvement in performance over single-contrast self-supervised methods. Single-contrast learned partitioning improves over single-contrast self-supervised by learning a partitioning distribution.

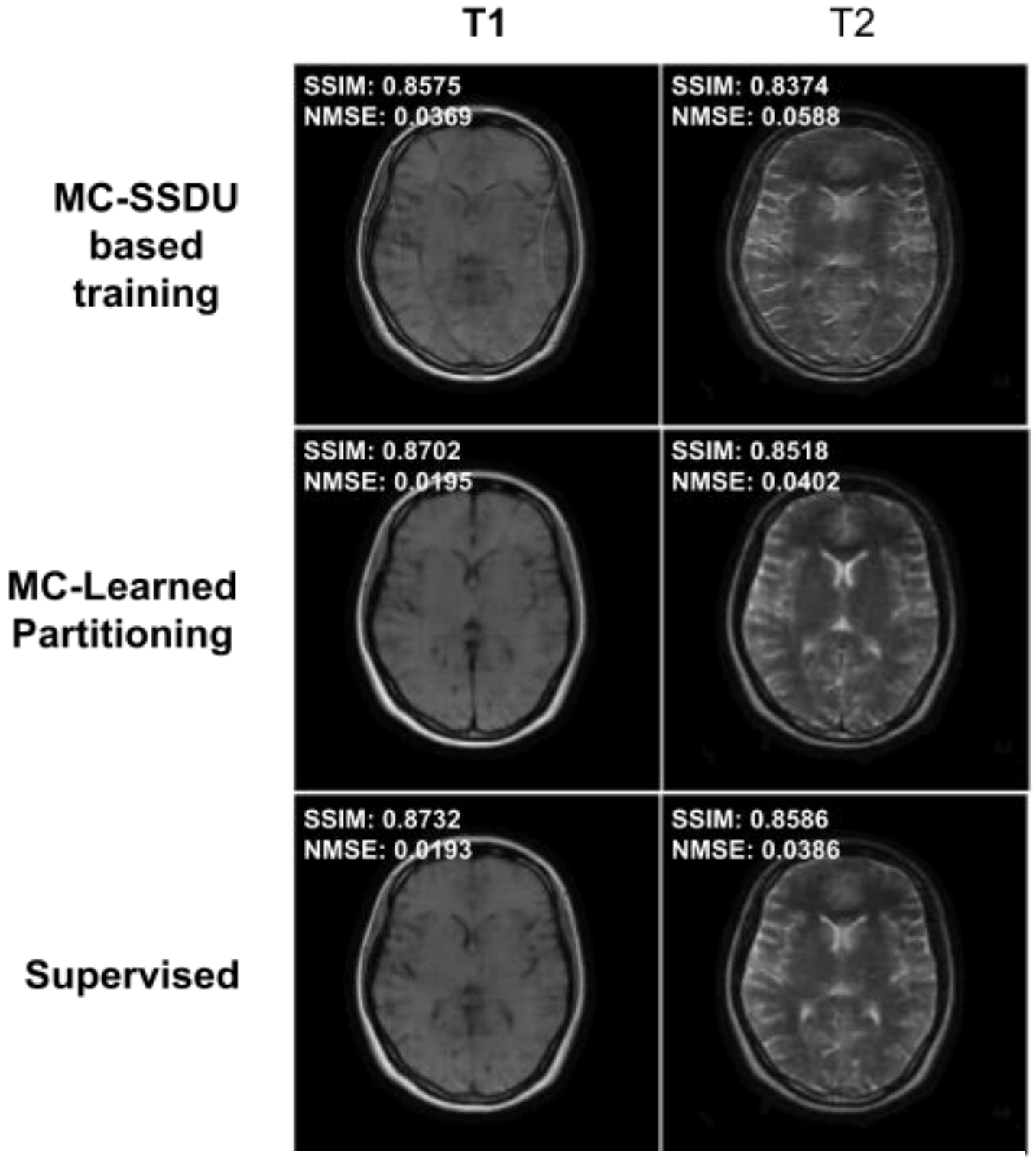

*FIG. 8.* Representative M4Raw slice (R=6, equal spaced parallel imaging pattern) reconstructed with IWNeXt.

### *F. Reconstruction Model Generalizability*

To assess the generalizability of our self-supervised training framework, we test on a different reconstruction network. Specifically, we use the IWNeXt multi-contrast network used to reconstruct one under-sampled contrast with a fully sampled guide contrast [49]. We modify the IWNeXt to accept multi-channel data by using the sensitivity estimation module and adapt the convolution weights to simultaneously reconstruct multiple contrasts. We compare fully supervised IWNeXt, IWNeXt trained with the SSDU self-supervised framework, and IWNeXt trained with the proposed dual-domain learned partitioning framework to evaluate whether the benefits of our training strategy generalize across reconstruction architectures.

Figure 8 shows the representative slices with the IWNeXt model. We use the same training setup for the M4Raw data and use an initial under-sampling pattern of equal spaced parallel imaging with R=6. Our method, MC-Learned partitioning, reaches the most similar performance to multi-contrast supervised.

## IV. Discussion

We introduce a multi-contrast learned partitioning framework to learn the optimal partitioning of data for SSDU which improves self-supervised learning performance. We additionally show that the multi-contrast self-supervised training framework can significantly improve performance over single-contrast self-supervised learning training regimes. Our multi-contrast self-supervised method performs better than single-contrast self-supervised learning at all acceleration factors and under-sampling patterns. This improved multi-contrast performance shows the benefit of different contrasts providing complementary information in reconstruction. This multi-contrast performance benefit can be supplemented with learned partitioning between contrasts, which further improves performance for many acceleration factors and sampling patterns.

At high acceleration factors (R=6, 8) for the 1D variable sampling pattern, our multi-contrast self-supervised framework performs better than single contrast supervised which has access to fully sampled data. One possible reason for improved performance over supervised training is that at high acceleration factors for the 1D variable density, only a small number of k-space lines are acquired. This small number of lines does not form a representative set of k-space, which is difficult for the model to fill in missing k-space values. However, in multi-contrast reconstruction the aggregate k-space coverage over contrasts is larger, albeit the larger k-space coverage comes from different contrasts. This greater coverage of k-space may be why the network can make better predictions, because the network may implicitly learn contrast transfer mechanisms. Testing higher acceleration factors of 2D variable density and equal-spaced may show the same trend, because less k-space data would be acquired.

The learned partitioning maps somewhat resemble the initial under-sampling distribution. The center region of k-space, the learned partitioning pattern resembles a star shaped pattern. One possible reason for this pattern is for improved sharpness of edges in the reconstruction output. Additionally, the 1D and 2D sampling pattern contains low probability regions, which may be optimal regions to calculate the loss over. We have also tested possible learned complementary sampling between contrasts, because previous work has shown that segregated sampling over multiple contrasts provides better performance [4], [18], [27]. However, we did not find any evidence of segregated sampling in our learned partitioning distributions. One reason for the lack of segregated sampling in our methods is that the model samples from the learned distributions from all contrasts independently. It is possible that improved performance can be achieved by more explicitly enforcing more segregated partitioning. This is left for future work.

We also show that there is improved performance as we increase the number of contrasts in reconstruction. We propose two possible reasons. First, by adding more contrasts, we get more of an averaging effect as the number of scans in our sequence increases. Because the network could act as a contrast transfer mechanism, each contrast in our multi-contrast reconstruction framework could be de-noised by the other contrasts. Second, as more contrasts are added, more k-space samples accumulate in the aggregate over contrasts. The network can then use contrast transfer mechanisms to use the aggregate multi-contrast k-space to better infer the missing values.

We show that these multi-contrast results hold even in retrospectively under-sampled real acquired k-space from the M4Raw dataset. Of note, the M4Raw dataset is unregistered, which [17] shows that a spatial shift of a few voxels between contrasts may be negligible to performance. Improved performance may be realized if we incorporate a spatial alignment module as in [50] for larger differences between contrasts. Using a spatial alignment module will be beneficial in future studies, because real acquired data will inherently contain motion between contrasts.

Our method comes with a few limitations. In [42], the authors theoretically justify SSDU by relating this process to a similar self-supervised denoising task. However, their justification did not require a disjoint partitioning of k-space, which was typically done in SSDU. Learning a non-disjoint partitioning of k-space similar to early works in self-supervised learning [51] may further improve the learned partitioning method, as more sampled k-space will be passed through the network.

Another notable limitation is the performance benefit of the learned partitioning method. In sequences where the optimal theoretical portioning is known (2D and 1D variable density) the benefit from the learned partitioning is limited. However, in the equal-spaced undersampling pattern, the benefit is greater due to no theoretical optimal partitioning distribution. The learned partitioning method may provide a grater performance benefit when incorporating self-supervised and multi-contrast denoising [52], [53], [54].

In Figure 8, we show our self-supervised partitioning framework can be used for a more advanced reconstruction network which includes attention. Multiple multi-contrast networks have been proposed specifically for multi-contrast reconstruction. Notably in [55] they find that a U-Net is a sub-optimal network choice for joint multi-contrast reconstruction. Instead, they use a modified U-Net with separate encoders and decoders for each contrast. Additional methods with transformers and multi-head attention have also been shown to further improve multi-contrast reconstruction performance [21], [22], [33].

## V. Conclusion

Here we propose a method that improves self-supervised performance by the introduction of multi-contrast information coupled with learning the partitioning distribution. These multi-contrast networks perform better to their single contrast counterparts, highlighting the complementary information sharing between contrasts. Additionally, the learned partitioning distribution resembles the initial under-sampling distribution $M_{\Omega}$ with a star shaped pattern near the center of k-space.

We show that these methods work for under-sampled simulated or real acquired low-field data. These results may not be indicative of true performance, because retrospectively under-sampled k-space overinflates performance compared to prospectively under-sampling [56]. Future work will directly evaluate these methods on prospectively under-sampled multi-contrast protocols of the same resolution.